\documentclass[a4paper,11pt]{article}
\usepackage{pos}

\newcommand{\eps}{\varepsilon}
\newcommand{\bs}[1]{\boldsymbol{#1}}

\title{Intersection theory and canonical differential equations}
\author[a]{Claude Duhr}
\author[a]{Sara Maggio}
\author[a]{Franziska Porkert}
\author[a]{Cathrin Semper}
\author[b]{Yoann Sohnle}
\author*[a]{Sven F. Stawinski}

\affiliation[a]{Bethe Center for Theoretical Physics, University of Bonn,\\
  Wegelerstr.\ 10, D-53115 Bonn, Germany}

\affiliation[b]{Department of Physics and Astronomy, Uppsala University,\\
Box 516, 75120 Uppsala, Sweden}

\emailAdd{cduhr@uni-bonn.de}
\emailAdd{smaggio@uni-bonn.de}
\emailAdd{fporkert@uni-bonn.de}
\emailAdd{csemeper@uni-bonn.de}
\emailAdd{yoann.sohnle@physics.uu.se}
\emailAdd{sstawins@uni-bonn.de}

\abstract{In these proceedings we will review recent progress in applying ideas from the mathematical framework of twisted cohomology to the study of canonical differential equations for Feynman integrals. Firstly, we will show how the intersection matrix can shed some light on the nature of the canonical basis of a Feynman integral family, a concept still not fully understood in the general case. In particular we will show how the intersection matrix can detect hidden linear dependencies of the iterated integrals resulting from an $\eps$-factorized differential equation, which are difficult to find otherwise. Furthermore, we will explain how the intersection matrix can help in deriving (polynomial) relations between the transcendental functions occurring in the rotation to the canonical basis. This allows us to simplify the rotation, and furthermore leads to simplifications in the final result. The discussion we be kept as light as possible, focusing on a simple running example and deferring the technical details to the original publications.
}

\FullConference{17th International Symposium on Radiative Corrections: Applications of Quantum Field Theory to Phenomenology (RADCOR2025)\\
5-10 October 2025\\
Puri, India\\}

\begin{document}

\begin{flushright}
    BONN-TH/2026-03 \\
    UUITP-01/26
\end{flushright}

\maketitle

\section{Introduction}
\label{sec:Intro}
Feynman integrals are a cornerstone of perturbative quantum field theory and their evaluation is crucial for precision studies of particle and gravitational wave physics. As is well known, however, Feynman integrals are typically divergent and need to be regulated. The most common way to achieve this is dimensional regularization, where the space-time dimension $D=d-2\varepsilon$ is deformed from an integer value $d$ by a generic parameter $\varepsilon$. While this is very convenient for calculations, it has a very immediate consequence, namely it leads to multivalued integrands. Integrals of this type are precisely the object of study in the mathematical framework of twisted cohomology or intersection theory. Hence this framework constitutes a natural language to talk about Feynman integrals in dimensional regularization \cite{Mastrolia:2018uzb}.

It is thus natural to rephrase various aspects of Feynman integral calculus in this language, giving us a new viewpoint and allowing us to learn more about Feynman integrals. Indeed, this has been pursued since this connection has been made close to a decade ago. However, so far, most focus has been on integration-by-parts (IBP) relations, i.e., linear relations between Feynman integrals \cite{Mastrolia:2018uzb,Frellesvig:2019kgj,Frellesvig:2019uqt,Weinzierl:2020xyy,Frellesvig:2020qot, Chestnov:2022alh,Chestnov:2022xsy,Jiang:2023oyq,Brunello:2023rpq,Fontana:2023amt,Brunello:2024tqf,Lu:2024dsb}. These are clearly of high relevance to practical calculations, as they allow to reduce any integral of a Feynman integral family to a finite set of \emph{master integrals}. However, it is also interesting to consider other aspects of Feynman integrals in this framework, see e.g., \cite{Mizera:2019vvs,Caron-Huot:2021xqj,Caron-Huot:2021iev,Giroux:2022wav,Duhr:2023bku,Chen:2023kgw,Duhr:2024rxe}. The focus of these proceedings will be the differential equations satisfied by the Feynman integrals \cite{Kotikov:1990kg,Kotikov:1991hm,Kotikov:1991pm,Gehrmann:1999as} and in particular their canonical bases \cite{Henn:2013pwa}. To this end we will follow \cite{Duhr:2024xsy,Duhr:2024uid,Duhr:2025xyy}, see also \cite{Chen:2020uyk,Chen:2022lzr} for related works. 

Explicitly, a chosen basis of master integrals $\bs{I}(\bs{x},\eps)$ satisfies a first-order differential equation
\begin{equation}
\label{eq:MIDE}
    \mathrm{d}\bs{I}(\bs{x},\varepsilon)=\bs{\Omega}(\bs{x},\eps)\bs{I}(\bs{x},\eps) \,,
\end{equation}
where the total derivative is taken with respect to the kinematic variables $\bs{x}=(x_1,\dots ,x_r)$ and $\bs{\Omega}(\bs{x},\eps)$ is a matrix of rational one-forms. The choice of master integrals is of course arbitrary and choosing a new basis $\bs{J}(\bs{x},\eps)=\bs{U}(\bs{x},\eps)\bs{I}(\bs{x},\eps)$, for some invertible matrix $\bs{U}(\bs{x},\eps)$, will lead a different form of the differential equation, related to the previous one by a non-abelian gauge transformation. A particularly convenient choice of basis is given by a \emph{canonical} or an \emph{$\varepsilon$-factorized basis} $\bs{J}(\bs{x},\eps)$, where the differential equation takes the form
\begin{equation}
    \mathrm{d}\bs{J}(\bs{x},\eps)=\eps\bs{A}(\bs{x})\bs{J}(\bs{x},\eps) \,,
\end{equation}
with $\bs{A}(\bs{x})$ independent of $\eps$. In the simplest cases, the canonical differential equation matrix $\bs{A}(\bs{x})$ additionally takes the form of a matrix of $\mathrm{dlog}$ forms of algebraic functions of the kinematical variables \cite{Henn:2013pwa}. In such cases it is also well understood how to construct the canonical basis, at least conceptually.

However, it is by now well-known that there are integrals which do not admit a canonical basis in such a $\mathrm{dlog}$ form, see \cite{Bourjaily:2022bwx} for a review. Indeed, there are integrals which admit an $\varepsilon$-factorized basis, but the rotation to this basis necessarily introduces transcendental functions into the differential equation that can not be written as $\mathrm{dlog}$s of algebraic arguments. However, even in these cases, $\varepsilon$-factorized bases still seem to exist very generally. Their properties are however less clear than in the $\mathrm{dlog}$ case. In particular there are $\varepsilon$-factorized bases with genuinely different properties and hence one needs to impose extra conditions to single out a ``canonical`` basis \cite{Frellesvig:2023iwr}. What exactly these additional requirements should be is however an open question. For the purpose of these proceedings we will adopt the definition of \cite{Duhr:2025lbz}, which requires a canonical basis to satisfy
\begin{enumerate}
    \item The basis is $\eps$-factorized.
    \item The entries of the differential equation matrix admit at most simple poles locally around each singularity.
    \item The linearly independent differential forms in the differential equation matrix are also linearly independent as cohomology classes, i.e., linearly independent up to total derivatives. For a more precise definition see \cite{Duhr:2024xsy}.
\end{enumerate}
In the rest of these proceedings we will study such canonical differential equations from the point of view of twisted cohomology. In particular we will explain how methods from this framework allow us to distinguish different $\eps$-factorized bases and to simplify the rotation to the canonical basis.
\section{The intersection matrix}
\label{sec:IntMatrix}
The object that will play the most important role in the following is the \emph{(twisted cohomology) intersection matrix} $\bs{C}(\bs{x},\eps)$. For details on its definition and methods for its computation see \cite{Mizera:2019gea,intNumbers,Frellesvig:2019kgj,Frellesvig:2019uqt,Weinzierl:2020xyy,Frellesvig:2020qot,Chestnov:2022xsy,Brunello:2023rpq,Fontana:2023amt,Brunello:2024tqf}. For our purposes we can simply think about it as a matrix of rational functions, arising from a bilinear pairing between ``Feynman integrands`` (typically thought of in a parametric representation) and ``\emph{dual} Feynman integrands``. We will not precisely define what dual Feynman integrands are (see e.g., \cite{Caron-Huot:2021iev,Caron-Huot:2021xqj}) but just note that throughout these proceedings we will focus on maximally cut Feynman integrals, for which one can view the dual Feynman integrands simply as the original Feynman integrands with the replacement $\eps\rightarrow -\eps$ \cite{Duhr:2024rxe}. This situation is referred to as \emph{self-duality} \cite{Duhr:2024xsy} and we will assume it in the following. In particular we can associate the intersection matrix to a chosen basis of master integrals $\bs{I}(\bs{x},\eps)$.\footnote{This statement is more subtle in the presence of symmetry relations between master integrals, see \cite{symmPaper}. This will however play no role in the following.}

Let us now record some properties of this matrix that will turn out to be important in the following. First of all the intersection matrix satisfies a first-order differential equation
\begin{equation}
\label{eq:CDE}
    \mathrm{d}\bs{C}(\bs{x},\eps)=\bs{\Omega}(\bs{x},\eps)\bs{C}(\bs{x},\eps)+\bs{C}(\bs{x},\eps)\bs{\Omega}(\bs{x},-\eps)^T \,,
\end{equation}
where $\bs{\Omega}(x,\eps)$ is the differential equation matrix in the chosen master integral basis, as in \eqref{eq:MIDE}. Furthermore, the intersection matrix appears in the so-called \emph{twisted Riemann bilinear relations} which we can write as\footnote{The twisted Riemann bilinear relations are really a matrix of equations, see e.g., \cite{Duhr:2024rxe}. The equation given here is obtained by projecting onto appropriate rows and columns.}
\begin{equation}
\label{eq:TRBR}
    \bs{I}(\bs{x},\eps)^T\bs{C}(\bs{x},\eps)^{-1T}\bs{I}(\bs{x},-\eps)=h(\eps) \,.
\end{equation}
Here the right-hand side is a scalar function, independent of the kinematics.
It follows from these relations that if the chosen master integral basis is a canonical basis in the sense described in Section \ref{sec:Intro} (the linear independence property is crucial here), then
\begin{equation}
\label{eq:CConst}
    \mathrm{d}\bs{C}(\bs{x},\eps)=0 \,,
\end{equation}
i.e., the intersection matrix is independent of the kinematics. The intuitive reason for this is the following. The linear independence property of the entries of the canonical differential equation implies linear independence of the iterated integrals appearing in the path-ordered exponential, furnishing the solution to the differential equation in the canonical basis \cite{iteratedIntIndependece,Duhr:2024xsy}. After expanding in $\eps$, the relation \eqref{eq:TRBR} however seems to imply such a linear relation, which hence has to be trivial, which is only possible if the intersection matrix does not depend on the kinematical variables. A rigorous proof of this statement was presented in \cite{Duhr:2024xsy}. 

Since we are here assuming self-duality (since we are focusing on maximal cuts), we can say even more about the intersection matrix. In particular, also the dependence on $\eps$ takes a simple form, namely it factorizes out into a scalar function $f(\eps)$,
\begin{equation}
\label{eq:deltaDef}
    \bs{C}(\bs{x},\eps)=f(\eps)\bs{\Delta} \,, \qquad \bs{\Delta}^T=\pm \bs{\Delta} \,.
\end{equation}
The matrix $\bs{\Delta}$, called the \emph{canonical intersection matrix}, is a purely numerical matrix, often a matrix of rational numbers. This matrix will play an important role in the following. The first application of this matrix is the fact that it encodes a symmetry of the canonical differential equation matrix
\begin{equation}
    \bs{\Delta}\bs{A}^T\bs{\Delta}^{-1}=\bs{A} \,,
\end{equation}
which immediately follows from \eqref{eq:CDE} in the canonical basis \cite{Duhr:2024xsy}. Such symmetries have been observed in examples before in \cite{Pogel:2024sdi}. In the following we will elucidate how the (canonical) intersection matrix can teach us more about the canonical basis.

\section{Detecting the canonical basis}
To keep the discussion as light as possible, we will consider an illustrative running example, given by the following integral family
\begin{equation}
    I_{\bs{\nu}}(x)=\int_0^1\mathrm{d}z\, z^{-\frac{1}{2}+\nu_1+a_1\eps}(1-z)^{-\frac{1}{2}+\nu_2+a_2\eps}(x-z)^{-\frac{1}{2}+\nu_3+a_3\eps} \,,
\end{equation}
where $\bs{\nu}=(\nu_1,\nu_2,\nu_3)$ is a vector of integers and $a_1,a_2,a_3$ are some rational numbers.\footnote{We will put $a_i=1$ in some expressions for brevity.} Furthermore, $\eps$ is a generic parameter and $x>1$ is taken to be real, for simplicity. Note that for $\eps=0$, this integral computes a period of the Legendre elliptic curve defined by the polynomial equation
\begin{equation}
    y^2=z(z-1)(z-x) \,.
\end{equation}
In this sense the above family acts as a toy model for the maximal cuts of an elliptic Feynman integral family. The parameter $\eps$ here plays the role of the dimensional regulator and the variable $x$ plays the role of the kinematics. This integral family admits two master integrals which we can for example choose to be
\begin{equation}
    \bs{I}=(I_{0,0,0},\, I_{1,0,0}) \,.
\end{equation}
They satisfy a differential equation as in \eqref{eq:MIDE}, with
\begin{equation}
\label{eq:OmegaMatrix}
    \bs{\Omega}(x,\eps)=
    \left[\frac{1}{2}\begin{pmatrix}
        -\frac{1}{x-1} & \frac{1}{x(x-1)} \\
        -\frac{1}{x-1} & \frac{1}{x-1}
    \end{pmatrix} 
    +\varepsilon\begin{pmatrix}
        -\frac{a_1}{x(x-1)}+\frac{a_3}{x} & \frac{s_1}{x(x-1)} \\
        -\frac{a_1}{x-1} & \frac{a_1+a_2+a_3}{x-1}
    \end{pmatrix}
    \right]\mathrm{d}x \,,
\end{equation}
Our first goal is to show how the intersection matrix can shed light on the distinction between different $\eps$-factorized bases. To this end observe that there are (at least) two ways of $\eps$-factorizing the differential equation \eqref{eq:MIDE} \cite{Frellesvig:2023iwr}. The first one, following \cite{Frellesvig:2021hkr}, can be obtained by noticing that the differential equation matrix $\bs{\Omega}(x,\eps)$ is linear in $\eps$. We can hence achieve $\eps$-factorization by dividing by the solution to the $\eps^0$ part of the differential equation, which is simply the period matrix of the elliptic curve
\begin{equation}
    \bs{W}=\begin{pmatrix}
        \omega_1(x) & \omega_2(x) \\ \eta_1(x) & \eta_2(x) 
    \end{pmatrix} \,.
\end{equation}
Here $\omega_1,\omega_2$ are the periods and $\eta_1,\eta_2$ are the quasi-periods of the Legendre curve and we can view them as being defined by the first order differential equation that they satisfy.\footnote{For definiteness we will in the following choose $\omega_1$ to be a holomorphic solution.} Explicitly, we find that the basis $\bs{K}(x,\eps)=\bs{W}^{-1}(x)\bs{I}(x,\eps)$ satisfies an $\eps$-factorized differential equation
\begin{equation}
    \mathrm{d}\bs{K}(x,\eps)=\eps\bs{B}(x)\bs{K}(x,\eps) \,.
\end{equation}
The concrete form of the matrix $\bs{B}(x)$ will not be important in the following, but to get an idea of the form of the entries let us simply record one of them here
\begin{equation}
\label{eq:B21entry}
    B_{21}(x)=\frac{a_1(\eta_1-\omega_1)(\eta_1-x\omega_1)+a_2\eta_1(\eta_1-x\omega_1)+a_3\eta_1(\eta_1-\omega_1)}{8\pi ix(x-1)}\mathrm{d}x \,,
\end{equation}
Note that we are suppressing the dependence of the (quasi-)periods on $x$ for legibility. 

Another way to $\eps$-factorize this differential equation is to follow the strategy outlined in \cite{Gorges:2023zgv,Duhr:2025lbz}. This has been carried out in detail in \cite{Duhr:2024uid} (see also \cite{Broedel:2018rwm}) and the result is the basis $\bs{J}(x,\eps)=\bs{U}(x,\eps)\bs{I}(x,\eps)$ 
with
\begin{equation}
\label{eq:URot}
    \bs{U}(x,\eps)=
    \begin{pmatrix}
        1 & 0 \\ t(x) & 1
    \end{pmatrix}
    \begin{pmatrix}
        \eps & 0 \\ 0 & 1
    \end{pmatrix}
    \begin{pmatrix}
        \frac{1}{\omega_1} & 0 \\ 
        \frac{\eta_1-x\omega_1}{8\pi i} & -\frac{x(x-1)\omega_1}{4\pi i}
    \end{pmatrix}
    \begin{pmatrix}
        1 & 0 \\
        -\frac{x+2a_1\eps-2a_3\eps(x-1)}{2x(x-1)} & \frac{1+2(a_1+a_2+a_3)\eps}{2x(x-1)}
    \end{pmatrix} \,,
\end{equation}
where we defined
\begin{equation}
\label{eq:tFct}
    t(x)=\frac{(a_1(x-1)+a_2x+a_3(2x-1))\omega_1^2}{8\pi i} \,.
\end{equation}
Also the basis $\bs{J}(x,\eps)$ satisfies an $\eps$-factorized differential equation $\mathrm{d}\bs{J}(x,\eps)=\eps\bs{A}(x)\bs{J}(x,\eps)$
% , which satisfies a differential equation $\mathrm{d}\bs{J}(x,\eps)=\eps\bs{A}(x)\bs{J}(x,\eps)$ 
with
\begin{equation}
\label{eq:AMat}
    \bs{A}(x)=
    \begin{pmatrix}
        \frac{2x-1}{x(x-1)} & -\frac{4\pi i}{x(x-1)\omega_1^2} \\
        -\frac{(x^2-x+1)\omega_1^2}{4\pi ix(x-1)} & \frac{2x-1}{x(x-1)}
    \end{pmatrix}\mathrm{d}x \,,
\end{equation}
where we put $a_i=1$ for brevity. While both bases are $\eps$-factorized and locally have at most simple poles, an important feature distinguishing these two bases is the property of linear independence, the third point in our definition of canonical, see Section \ref{sec:Intro}. Indeed, while the single entry given in \eqref{eq:B21entry} is already a total derivative by itself
\begin{equation}
    B_{21}(x)=\mathrm{d}\left(
    \frac{\omega_1(a_1(2\eta_1-(x+1)\omega_1)+a_2(2\eta_1-x\omega_1)+a_3(2\eta_1-\omega_1))}{8\pi i}
    \right) \,,
\end{equation}
the linearly independent entries of \eqref{eq:AMat} are also linearly independent as cohomology classes, which is not difficult to prove using the same arguments as in \cite{Duhr:2025xyy}. In general, it is however not easy to see whether there is such a linear relation. There is however a simple check one can do, which is provided by the criterion \eqref{eq:CConst}. Indeed, if the intersection matrix is not constant in the $\eps$-factorized basis, then there has to be a linear relation and the basis can not be canonical in the sense of Section \ref{sec:Intro}. The converse direction is currently still an open question. In our concrete example we indeed find that the intersection matrix $\bs{C}_K(x,\eps)$ in the basis $\bs{K}(x,\eps)$ is not constant (the concrete form is not important), while in the basis $\bs{J}(x,\eps)$ we find
\begin{equation}
    \bs{C}_J(x,\eps)=\frac{\eps}{8\pi^2}\begin{pmatrix}
        0 & 1 \\ 1 & 0
    \end{pmatrix} \,,
\end{equation}
in accordance with our expectations from Section \ref{sec:IntMatrix}.
\section{Simplifying the canonical rotation}
The second application of the intersection matrix we want to showcase here, is that it can help in simplifying the rotation to the canonical basis and as such also the canonical differential equation and hence the result. More precisely we will focus on the last step of the algorithm of \cite{Gorges:2023zgv,Duhr:2025lbz} (it can in the same way be applied to the method of \cite{e-collaboration:2025frv}). In our running example this means that we perform all but the last rotation in \eqref{eq:URot}. This leads to a differential equation of the form
\begin{equation}
    \bs{\Omega}'(x,\eps)=
    \left[\begin{pmatrix}
        0 & 0 \\ p(x) & 0 
    \end{pmatrix}
    +\eps
    \begin{pmatrix}
        0 & -\frac{4\pi i}{x(x-1)\omega_1^2} \\
        \frac{3\omega_1^2}{4\pi i} & \frac{2(2x-1)}{x(x-1)}
    \end{pmatrix}\right]\mathrm{d}x \,, \qquad p(x)=\frac{\omega_1(x\omega_1+(1-2x)\eta_1)}{4\pi ix(x-1)} \,,
\end{equation}
again putting $a_i=1$, where the only non $\eps$-factorized term is below the diagonal. The final rotation in \eqref{eq:URot} then introduces a function $t(x)$, which satisfies a first-order differential equation $dt(x)=p(x)\mathrm{d}x$ such that it cancels this term. For this reason such functions were called \emph{$\eps$-functions} in \cite{Duhr:2025kkq,Duhr:2025xyy}.
In this case it is easy to see that the differential equation for the $\eps$-function $t(x)$ can be integrated up in terms of the period $\omega_1$, leading to \eqref{eq:tFct}. In general this may however become a difficult problem. 

However, as we will see now, we can also determine $t(x)$ without solving any differential equation, by just using the requirement of the constancy of the intersection matrix \eqref{eq:CConst}. The idea, first proposed in \cite{Duhr:2024uid}, is to compute the intersection matrix in the original basis
\begin{equation}
    \bs{C}_I(x,\eps)=\begin{pmatrix}
        0 & \frac{1}{i\pi(1-2s_1\eps)} \\
       -\frac{1}{i\pi(1+2s_1\eps)} & \frac{2\eps((1+x)a_1+xa_2+a_3)}{i\pi(1-2s_1\eps)(1+2s_1\eps)}
    \end{pmatrix} \,,
\end{equation}
and apply the rotation \eqref{eq:URot} to the basis $\bs{J}(x,\eps)$, leaving the function $t(x)$ unspecified
\begin{equation}
    \bs{C}_J(x,\eps)=\bs{U}(x,\eps)\bs{C}_I(x,\eps)\bs{U}(x,-\eps)^T=\frac{\eps}{8\pi^2}
    \begin{pmatrix}
        0 & 1 \\ 1 & 2t(x)-\frac{((x-1)a_1+xa_2+(2x-1)a_3)\omega_1^2}{4\pi i}
    \end{pmatrix} \,.
\end{equation}
The requirement \eqref{eq:CConst} hence determines the function $t(x)$ to be precisely \eqref{eq:tFct}, up to an irrelevant integration constant.

This idea generalizes nicely to more complicated situations. In general, the final rotation also depends on $\eps$ and takes the form of a lower uni-triangular matrix that we can expand as
\begin{equation}
    \bs{U}_{\mathrm{t}}(\bs{x},\eps)=\frac{1}{\eps^{n-1}}\bs{U}^{(-(n-1))}_{\mathrm{t}}(\bs{x},\eps)+\dots + \bs{U}^{(0)}_{\mathrm{t}}(\bs{x},\eps) \,.
\end{equation}
Here the non-trivial entries of $\bs{U}^{(0)}_{\mathrm{t}}(\bs{x},\eps)$ are given by some $\eps$-functions $t_{ij}(\bs{x})$, while the lower $\eps$-orders generally contain derivatives of these, which however do not contribute to the final differential equation. The $\eps$-functions are defined by a system of coupled partial differential equations in such a way that they cancel the remaining non $\eps$-factorized terms below the diagonal. In general these differential equations can be integrated up in terms of (iterated) integrals of algebraic functions and (quasi-)periods of some algebraic varieties but it is in general not clear if and how many linear combinations of these can be written in terms of simpler functions, as was possible in our simple example.

Following the same logic as above, we can however derive polynomial relations between the $\eps$-functions relating (some of) them to simpler objects. The general picture is the following. As seen in \eqref{eq:deltaDef}, the intersection matrix gives rise to a numerical matrix $\bs{\Delta}$, the canonical intersection matrix. We use this matrix to split up the $\eps^0$ part of the final rotation into an \emph{orthogonal} and a \emph{symmetric} part
\begin{equation}
    \bs{U}^{(0)}_{\mathrm{t}}(\bs{x},\eps)=\bs{O}(\bs{x},\eps)\bs{R}(\bs{x},\eps) \,,
\end{equation}
where orthogonal and symmetric are understood in a generalized sense, i.e.,
\begin{equation}
    \bs{\Delta}\bs{O}(\bs{x},\eps)^T\bs{\Delta}^{-1}=\bs{O}(\bs{x},\eps)^{-1}, \qquad \bs{\Delta}\bs{R}(\bs{x},\eps)^T\bs{\Delta}^{-1}=\bs{R}(\bs{x},\eps) \,.
\end{equation}
Then, the elements of the symmetric part are precisely the combinations of $\eps$-functions that are fixed in terms of simpler objects, while the elements of the orthogonal part are not fixed by these considerations. This splitting hence gives an upper bound on the number of new transcendental objects that one needs to introduce to find the canonical basis.
Indeed, in examples it is possible to prove that this upper bound is also optimal and we expect this to be true more generally \cite{Duhr:2025xyy}. This method has by now been applied to maximal cuts of Feynman integrals associated with various non-trivial varieties, such as Calabi-Yau manifolds \cite{Duhr:2025xyy} or higher-genus Riemann surfaces \cite{Duhr:2024uid}. It has also been applied in the computation of the four-mass three-loop banana integral family (see also \cite{Pogel:2025bca}), where it has allowed remarkable simplifications of the resulting differential equation \cite{Duhr:2025kkq}.

\section{Conclusions}
It is by now well understood that twisted cohomology provides a natural framework to study Feynman integrals in dimensional regularization. In these proceedings we have reviewed how recasting the concept of canonical differential equations in the language of twisted cohomology can provide both conceptual as well as practical input. The key player here has been the (canonical) intersection matrix. Firstly, we have argued that the intersection matrix can ``detect`` linear dependencies (between cohomology classes) in an $\eps$-factorized differential equation. This can help in distinguishing different $\eps$-factorized bases since finding such linear relation can be difficult in practice. Furthermore, we have shown how the canonical intersection matrix allows us to derive relations between the $\eps$-functions that one needs to introduce to derive the canonical form. This sheds some light on these, still poorly understood, functions and can help in simplifying the results of such computations. In the future it will be interesting to further develop the links between twisted cohomology and Feynman integrals, as there are certainly more lessons to be learned. 

\acknowledgments

The work of CS is supported by the CRC 1639 ``NuMeriQS'', and the work of CD, FP, SM and SFS is funded by the European Union
(ERC Consolidator Grant LoCoMotive 101043686). Views
and opinions expressed are however those of the author(s)
only and do not necessarily reflect those of the European
Union or the European Research Council. Neither the
European Union nor the granting authority can be held
responsible for them. 
YS acknowledges support from the Centre for Interdisciplinary Mathematics at Uppsala University and partial support by the European Research Council under ERC Synergy Grant MaScAmp
101167287.

\bibliographystyle{JHEP}
\bibliography{References}

@inproceedings{Broedel:2018rwm,
    author = "Broedel, Johannes and Duhr, Claude and Dulat, Falko and Penante, Brenda and Tancredi, Lorenzo",
    title = "{From modular forms to differential equations for Feynman integrals}",
    booktitle = "{KMPB Conference}: {Elliptic Integrals, Elliptic Functions and Modular  Forms in Quantum Field Theory}",
    eprint = "1807.00842",
    archivePrefix = "arXiv",
    primaryClass = "hep-th",
    doi = "10.1007/978-3-030-04480-0_6",
    pages = "107--131",
    year = "2019"
}

@article{Duhr:2025lbz,
    author = "Duhr, Claude and Maggio, Sara and Nega, Christoph and Sauer, Benjamin and Tancredi, Lorenzo and Wagner, Fabian J.",
    title = "{Aspects of canonical differential equations for Calabi-Yau geometries and beyond}",
    eprint = "2503.20655",
    archivePrefix = "arXiv",
    primaryClass = "hep-th",
    reportNumber = "BONN-TH-2025-11, TUM-HEP 1559/25, HU-EP-25/13-RTG",
    doi = "10.1007/JHEP06(2025)128",
    journal = "JHEP",
    volume = "06",
    pages = "128",
    year = "2025"
}

@article{Duhr:2024xsy,
    author = "Duhr, Claude and Porkert, Franziska and Semper, Cathrin and Stawinski, Sven F.",
    title = "{Self-duality from twisted cohomology}",
    eprint = "2408.04904",
    archivePrefix = "arXiv",
    primaryClass = "hep-th",
    reportNumber = "BONN-TH-2024-11",
    doi = "10.1007/JHEP03(2025)053",
    journal = "JHEP",
    volume = "03",
    pages = "053",
    year = "2025"
}

@article{Gorges:2023zgv,
    author = {G{\"o}rges, Lennard and Nega, Christoph and Tancredi, Lorenzo and Wagner, Fabian J.},
    title = "{On a procedure to derive {\ensuremath{\epsilon}}-factorised differential equations beyond polylogarithms}",
    eprint = "2305.14090",
    archivePrefix = "arXiv",
    primaryClass = "hep-th",
    doi = "10.1007/JHEP07(2023)206",
    journal = "JHEP",
    volume = "07",
    pages = "206",
    year = "2023"
}

@article{Duhr:2024uid,
    author = "Duhr, Claude and Porkert, Franziska and Stawinski, Sven F.",
    title = "{Canonical differential equations beyond genus one}",
    eprint = "2412.02300",
    archivePrefix = "arXiv",
    primaryClass = "hep-th",
    reportNumber = "BONN-TH-2024-17",
    doi = "10.1007/JHEP02(2025)014",
    journal = "JHEP",
    volume = "02",
    pages = "014",
    year = "2025"
}

@article{Frellesvig:2023iwr,
    author = "Frellesvig, Hjalte and Weinzierl, Stefan",
    title = "{On $\varepsilon$-factorised bases and pure Feynman integrals}",
    eprint = "2301.02264",
    archivePrefix = "arXiv",
    primaryClass = "hep-th",
    reportNumber = "MITP/23-001",
    doi = "10.21468/SciPostPhys.16.6.150",
    journal = "SciPost Phys.",
    volume = "16",
    number = "6",
    pages = "150",
    year = "2024"
}

@article{Frellesvig:2021hkr,
    author = "Frellesvig, Hjalte",
    title = "{On epsilon factorized differential equations for elliptic Feynman integrals}",
    eprint = "2110.07968",
    archivePrefix = "arXiv",
    primaryClass = "hep-th",
    doi = "10.1007/JHEP03(2022)079",
    journal = "JHEP",
    volume = "03",
    pages = "079",
    year = "2022"
}

@article{Duhr:2025xyy,
    author = "Duhr, Claude and Maggio, Sara and Porkert, Franziska and Semper, Cathrin and Sohnle, Yoann and Stawinski, Sven F.",
    title = "{Canonical differential equations and intersection matrices}",
    eprint = "2509.17787",
    archivePrefix = "arXiv",
    primaryClass = "hep-th",
    reportNumber = "BONN-TH/2025-30, UUITP--27/25",
    month = "9",
    year = "2025"
}

@article{e-collaboration:2025frv,
    author = "Bree, Iris and others",
    collaboration = "{\ensuremath{\varepsilon}}-collaboration",
    title = "{The geometric bookkeeping guide to Feynman integral reduction and $\varepsilon$-factorised differential equations}",
    eprint = "2506.09124",
    archivePrefix = "arXiv",
    primaryClass = "hep-th",
    month = "6",
    year = "2025"
}

@article{Mastrolia:2018uzb,
    author = "Mastrolia, Pierpaolo and Mizera, Sebastian",
    title = "{Feynman Integrals and Intersection Theory}",
    eprint = "1810.03818",
    archivePrefix = "arXiv",
    primaryClass = "hep-th",
    doi = "10.1007/JHEP02(2019)139",
    journal = "JHEP",
    volume = "02",
    pages = "139",
    year = "2019"
}

@article{Frellesvig:2019kgj,
    author = "Frellesvig, Hjalte and Gasparotto, Federico and Laporta, Stefano and Mandal, Manoj K. and Mastrolia, Pierpaolo and Mattiazzi, Luca and Mizera, Sebastian",
    title = "{Decomposition of Feynman Integrals on the Maximal Cut by Intersection Numbers}",
    eprint = "1901.11510",
    archivePrefix = "arXiv",
    primaryClass = "hep-ph",
    doi = "10.1007/JHEP05(2019)153",
    journal = "JHEP",
    volume = "05",
    pages = "153",
    year = "2019"
}

@article{Frellesvig:2019uqt,
    author = "Frellesvig, Hjalte and Gasparotto, Federico and Mandal, Manoj K. and Mastrolia, Pierpaolo and Mattiazzi, Luca and Mizera, Sebastian",
    title = "{Vector Space of Feynman Integrals and Multivariate Intersection Numbers}",
    eprint = "1907.02000",
    archivePrefix = "arXiv",
    primaryClass = "hep-th",
    doi = "10.1103/PhysRevLett.123.201602",
    journal = "Phys. Rev. Lett.",
    volume = "123",
    number = "20",
    pages = "201602",
    year = "2019"
}

@article{Frellesvig:2020qot,
    author = "Frellesvig, Hjalte and Gasparotto, Federico and Laporta, Stefano and Mandal, Manoj K. and Mastrolia, Pierpaolo and Mattiazzi, Luca and Mizera, Sebastian",
    title = "{Decomposition of Feynman Integrals by Multivariate Intersection Numbers}",
    eprint = "2008.04823",
    archivePrefix = "arXiv",
    primaryClass = "hep-th",
    doi = "10.1007/JHEP03(2021)027",
    journal = "JHEP",
    volume = "03",
    pages = "027",
    year = "2021"
}

@article{Chestnov:2022alh,
    author = "Chestnov, Vsevolod and Gasparotto, Federico and Mandal, Manoj K. and Mastrolia, Pierpaolo and Matsubara-Heo, Saiei J. and Munch, Henrik J. and Takayama, Nobuki",
    title = "{Macaulay matrix for Feynman integrals: linear relations and intersection numbers}",
    eprint = "2204.12983",
    archivePrefix = "arXiv",
    primaryClass = "hep-th",
    doi = "10.1007/JHEP09(2022)187",
    journal = "JHEP",
    volume = "09",
    pages = "187",
    year = "2022"
}

@article{Chestnov:2022xsy,
    author = "Chestnov, Vsevolod and Frellesvig, Hjalte and Gasparotto, Federico and Mandal, Manoj K. and Mastrolia, Pierpaolo",
    title = "{Intersection numbers from higher-order partial differential equations}",
    eprint = "2209.01997",
    archivePrefix = "arXiv",
    primaryClass = "hep-th",
    doi = "10.1007/JHEP06(2023)131",
    journal = "JHEP",
    volume = "06",
    pages = "131",
    year = "2023"
}

@article{Brunello:2023rpq,
    author = "Brunello, Giacomo and Chestnov, Vsevolod and Crisanti, Giulio and Frellesvig, Hjalte and Mandal, Manoj K. and Mastrolia, Pierpaolo",
    title = "{Intersection numbers, polynomial division and relative cohomology}",
    eprint = "2401.01897",
    archivePrefix = "arXiv",
    primaryClass = "hep-th",
    doi = "10.1007/JHEP09(2024)015",
    journal = "JHEP",
    volume = "09",
    pages = "015",
    year = "2024"
}

@article{Brunello:2024tqf,
    author = "Brunello, Giacomo and Chestnov, Vsevolod and Mastrolia, Pierpaolo",
    title = "{Intersection numbers from companion tensor algebra}",
    eprint = "2408.16668",
    archivePrefix = "arXiv",
    primaryClass = "hep-th",
    doi = "10.1007/JHEP07(2025)045",
    journal = "JHEP",
    volume = "07",
    pages = "045",
    year = "2025"
}

@article{Fontana:2023amt,
    author = "Fontana, Gaia and Peraro, Tiziano",
    title = "{Reduction to master integrals via intersection numbers and polynomial expansions}",
    eprint = "2304.14336",
    archivePrefix = "arXiv",
    primaryClass = "hep-ph",
    reportNumber = "ZU-TH 19/23",
    doi = "10.1007/JHEP08(2023)175",
    journal = "JHEP",
    volume = "08",
    pages = "175",
    year = "2023"
}

@article{Mizera:2019vvs,
    author = "Mizera, Sebastian and Pokraka, Andrzej",
    title = "{From Infinity to Four Dimensions: Higher Residue Pairings and Feynman Integrals}",
    eprint = "1910.11852",
    archivePrefix = "arXiv",
    primaryClass = "hep-th",
    doi = "10.1007/JHEP02(2020)159",
    journal = "JHEP",
    volume = "02",
    pages = "159",
    year = "2020"
}

@article{Caron-Huot:2021xqj,
    author = "Caron-Huot, Simon and Pokraka, Andrzej",
    title = "{Duals of Feynman integrals. Part I. Differential equations}",
    eprint = "2104.06898",
    archivePrefix = "arXiv",
    primaryClass = "hep-th",
    doi = "10.1007/JHEP12(2021)045",
    journal = "JHEP",
    volume = "12",
    pages = "045",
    year = "2021"
}

@article{Caron-Huot:2021iev,
    author = "Caron-Huot, Simon and Pokraka, Andrzej",
    title = "{Duals of Feynman Integrals. Part II. Generalized unitarity}",
    eprint = "2112.00055",
    archivePrefix = "arXiv",
    primaryClass = "hep-th",
    doi = "10.1007/JHEP04(2022)078",
    journal = "JHEP",
    volume = "04",
    pages = "078",
    year = "2022"
}

@article{Giroux:2022wav,
    author = "Giroux, Mathieu and Pokraka, Andrzej",
    title = "{Loop-by-loop differential equations for dual (elliptic) Feynman integrals}",
    eprint = "2210.09898",
    archivePrefix = "arXiv",
    primaryClass = "hep-th",
    doi = "10.1007/JHEP03(2023)155",
    journal = "JHEP",
    volume = "03",
    pages = "155",
    year = "2023"
}

@article{Chen:2020uyk,
    author = "Chen, Jiaqi and Jiang, Xuhang and Xu, Xiaofeng and Yang, Li Lin",
    title = "{Constructing canonical Feynman integrals with intersection theory}",
    eprint = "2008.03045",
    archivePrefix = "arXiv",
    primaryClass = "hep-th",
    doi = "10.1016/j.physletb.2021.136085",
    journal = "Phys. Lett. B",
    volume = "814",
    pages = "136085",
    year = "2021"
}

@article{Jiang:2023oyq,
    author = "Jiang, Xuhang and Lian, Ming and Yang, Li Lin",
    title = "{Recursive structure of Baikov representations: The top-down reduction with intersection theory}",
    eprint = "2312.03453",
    archivePrefix = "arXiv",
    primaryClass = "hep-ph",
    doi = "10.1103/PhysRevD.109.076020",
    journal = "Phys. Rev. D",
    volume = "109",
    number = "7",
    pages = "076020",
    year = "2024"
}

@article{Chen:2022lzr,
    author = "Chen, Jiaqi and Jiang, Xuhang and Ma, Chichuan and Xu, Xiaofeng and Yang, Li Lin",
    title = "{Baikov representations, intersection theory, and canonical Feynman integrals}",
    eprint = "2202.08127",
    archivePrefix = "arXiv",
    primaryClass = "hep-th",
    doi = "10.1007/JHEP07(2022)066",
    journal = "JHEP",
    volume = "07",
    pages = "066",
    year = "2022"
}

@article{Lu:2024dsb,
    author = "Lu, Mingming and Wang, Ziwen and Yang, Li Lin",
    title = "{Intersection theory, relative cohomology and the Feynman parametrization}",
    eprint = "2411.05226",
    archivePrefix = "arXiv",
    primaryClass = "hep-th",
    doi = "10.1007/JHEP05(2025)158",
    journal = "JHEP",
    volume = "05",
    pages = "158",
    year = "2025"
}

@article{Chen:2023kgw,
    author = "Chen, Jiaqi and Feng, Bo and Yang, Li Lin",
    title = "{Intersection theory rules symbology}",
    eprint = "2305.01283",
    archivePrefix = "arXiv",
    primaryClass = "hep-th",
    doi = "10.1007/s11433-023-2239-8",
    journal = "Sci. China Phys. Mech. Astron.",
    volume = "67",
    number = "2",
    pages = "221011",
    year = "2024"
}

@article{Duhr:2024rxe,
    author = "Duhr, Claude and Porkert, Franziska and Semper, Cathrin and Stawinski, Sven F.",
    title = "{Twisted Riemann bilinear relations and Feynman integrals}",
    eprint = "2407.17175",
    archivePrefix = "arXiv",
    primaryClass = "hep-th",
    reportNumber = "BONN-TH-2024-10",
    doi = "10.1007/JHEP03(2025)019",
    journal = "JHEP",
    volume = "03",
    pages = "019",
    year = "2025"
}

@article{Duhr:2025kkq,
    author = "Duhr, Claude and Maggio, Sara and Porkert, Franziska and Semper, Cathrin and Stawinski, Sven F.",
    title = "{Three-loop banana integrals with four unequal masses}",
    eprint = "2507.23061",
    archivePrefix = "arXiv",
    primaryClass = "hep-th",
    reportNumber = "BONN-TH-2025-24",
    doi = "10.1007/JHEP12(2025)034",
    journal = "JHEP",
    volume = "12",
    pages = "034",
    year = "2025"
}

@article{Pogel:2025bca,
    author = {P{\"o}gel, Sebastian and Teschke, Toni and Wang, Xing and Weinzierl, Stefan},
    title = "{The unequal-mass three-loop banana integral}",
    eprint = "2507.23594",
    archivePrefix = "arXiv",
    primaryClass = "hep-th",
    month = "7",
    year = "2025"
}

@article{Duhr:2023bku,
    author = "Duhr, Claude and Porkert, Franziska",
    title = "{Feynman integrals in two dimensions and single-valued hypergeometric functions}",
    eprint = "2309.12772",
    archivePrefix = "arXiv",
    primaryClass = "hep-th",
    reportNumber = "BONN-TH-2023-09",
    doi = "10.1007/JHEP02(2024)179",
    journal = "JHEP",
    volume = "02",
    pages = "179",
    year = "2024"
}

@article{Kotikov:1990kg,
	author = {Kotikov, A. V.},
	doi = {10.1016/0370-2693(91)90413-K},
	journal = {Phys. Lett. B},
	pages = {158--164},
	reportnumber = {ITF-90-31E},
	title = {{Differential equations method: New technique for massive Feynman diagrams calculation}},
	volume = {254},
	year = {1991}}

@article{Kotikov:1991hm,
	author = {Kotikov, A. V.},
	doi = {10.1016/0370-2693(91)90834-D},
	journal = {Phys. Lett. B},
	pages = {314--322},
	title = {{Differential equations method: The Calculation of vertex type Feynman diagrams}},
	volume = {259},
	year = {1991}}

@article{Kotikov:1991pm,
	author = {Kotikov, A. V.},
	doi = {10.1016/0370-2693(91)90536-Y},
	journal = {Phys. Lett. B},
	note = {[Erratum: Phys.Lett.B 295, 409--409 (1992)]},
	pages = {123--127},
	title = {{Differential equation method: The Calculation of N point Feynman diagrams}},
	volume = {267},
	year = {1991}}

@article{Gehrmann:1999as,
	archiveprefix = {arXiv},
	author = {Gehrmann, T. and Remiddi, E.},
	doi = {10.1016/S0550-3213(00)00223-6},
	eprint = {hep-ph/9912329},
	journal = {Nucl. Phys. B},
	pages = {485--518},
	reportnumber = {TTP-99-49},
	title = {{Differential equations for two-loop four-point functions}},
	volume = {580},
	year = {2000}}

@article{Henn:2013pwa,
	archiveprefix = {arXiv},
	author = {Henn, Johannes M.},
	doi = {10.1103/PhysRevLett.110.251601},
	eprint = {1304.1806},
	journal = {Phys. Rev. Lett.},
	pages = {251601},
	primaryclass = {hep-th},
	title = {{Multiloop integrals in dimensional regularization made simple}},
	volume = {110},
	year = {2013}}

@inproceedings{Bourjaily:2022bwx,
    author = "Bourjaily, Jacob L. and others",
    title = "{Functions Beyond Multiple Polylogarithms for Precision Collider Physics}",
    booktitle = "{Snowmass 2021}",
    eprint = "2203.07088",
    archivePrefix = "arXiv",
    primaryClass = "hep-ph",
    reportNumber = "BONN-TH-2022-05, UUITP-11/22, CERN-TH-2022-029, TUM-HEP-1391/22,
  HU-EP-22/08, MITP-22-022",
    year = "2022"
}

@phdthesis{Mizera:2019gea,
    author = "Mizera, Sebastian",
    title = "{Aspects of Scattering Amplitudes and Moduli Space Localization}",
    eprint = "1906.02099",
    archivePrefix = "arXiv",
    primaryClass = "hep-th",
    school = "Princeton, Inst. Advanced Study",
    year = "2020"
}

@article{intNumbers, 
    title={Intersection theory for twisted cohomologies and twisted Riemann’s period relations I}, 
    volume={139}, 
    DOI={10.1017/S0027763000005304}, 
    journal={Nagoya Math. J.}, 
    author={Cho, K. and Matsumoto, K.}, 
    year={1995}, 
    pages={67}}

@article{Weinzierl:2020xyy,
    author = "Weinzierl, Stefan",
    title = "{On the computation of intersection numbers for twisted cocycles}",
    eprint = "2002.01930",
    archivePrefix = "arXiv",
    primaryClass = "math-ph",
    doi = "10.1063/5.0054292",
    journal = "J. Math. Phys.",
    volume = "62",
    number = "7",
    pages = "072301",
    year = "2021"
}

@incollection{iteratedIntIndependece,
	author = {Deneufch{\^a}tel, Matthieu and Duchamp, G{\'e}rard H.E. and Hoang Ngoc Minh, Vincel and Solomon, Allan I.},
	booktitle = {{Lecture Notes in Computer Science, Volume 6742}},
	title = {{Independence of hyperlogarithms over function fields via algebraic combinatorics.}},
	url = {https://hal.science/hal-00558773},
	year = {2011}}

@article{Pogel:2024sdi,
    author = {P{\"o}gel, Sebastian and Wang, Xing and Weinzierl, Stefan and Wu, Konglong and Xu, Xiaofeng},
    title = "{Self-dualities and Galois symmetries in Feynman integrals}",
    eprint = "2407.08799",
    archivePrefix = "arXiv",
    primaryClass = "hep-th",
    reportNumber = "MITP/24-059, TUM-HEP-1513/24",
    doi = "10.1007/JHEP09(2024)084",
    journal = "JHEP",
    volume = "09",
    pages = "084",
    year = "2024"
}

@article{symmPaper,
    author = "Duhr, C. and Maggio, S. and Semper, C. and Stawinski, S.F.",
    title = "Symmetries of Feynman integrals and twisted cohomology",
    journal = "In preparation",
    year = "2026"
}

% \begin{thebibliography}{99}
% \bibitem{Duhr:2025lbz}
% C.\ Duhr, S.\ Maggio, C.\ Nega, B.\ Sauer, L.\ Tancredi F.J.\ Wagner, \emph{Aspects of canonical differential equations for Calabi-Yau geometries and beyond},
% \href{https://doi.org/10.1007/JHEP06(2025)128}
% {\emph{JHEP} \textbf{06} (2025) 128}
% [{\tt 2503.20655}].

% \bibitem{Duhr:2025lbz}
% C.\ Duhr, S.\ Maggio, C.\ Nega, B.\ Sauer, L.\ Tancredi F.J.\ Wagner, \emph{Aspects of canonical differential equations for Calabi-Yau geometries and beyond},
% \href{https://doi.org/10.1007/JHEP06(2025)128}
% {\emph{JHEP} \textbf{06} (2025) 128}
% [{\tt 2503.20655}].

% \bibitem{Duhr:2024xsy}
% C.\ Duhr, F.\ Porkert, C.\ Semper, S.F.\ Stawinski,
% \emph{Self-duality from twisted cohomology},
% \href{https://doi.org/10.1007/JHEP03(2025)053}
% {\emph{JHEP} \textbf{03} (2025) 053}
% [{\tt 2408.04904}].

% \end{thebibliography}

\end{document}